\documentclass[aps,12pt,a4paper,showpacs,byrevtex]{revtex4}
\usepackage{amsmath}
\usepackage{bm}
\usepackage{amsfonts}
\begin{document}
\baselineskip 15pt

\title{Time evolution, cyclic solutions and geometric phases for
general spin in an arbitrarily varying magnetic
field}\thanks{published in J. Phys. A {\textbf 36} (2003) 6799-6806.}



\author{Qiong-Gui Lin}
\email[]{qg_lin@163.net}


\affiliation{China Center of Advanced Science and Technology (World
Laboratory), P. O. Box 8730, Beijing 100080, People's Republic of
China}
\thanks{not for correspondence}
\affiliation{Department of Physics, Sun Yat-Sen University, Guangzhou
510275, People's  Republic of China}


\begin{abstract}
\baselineskip 15pt {\normalsize A neutral particle with general spin
and magnetic moment moving in an arbitrarily varying magnetic field
is studied. The time evolution operator for the Schr\"odinger
equation can be obtained if one can find a unit vector that satisfies
the equation obeyed by the mean of the spin operator. There exist at
least $2s+1$ cyclic solutions in any time interval. Some particular
time interval may exist in which all solutions are cyclic. The
nonadiabatic geometric phase for cyclic solutions generally contains
extra terms in addition to the familiar one that is proportional to
the solid angle subtended by the closed trace of the spin vector.}
\end{abstract}

\maketitle



Since the discovery of the geometric phase
\cite{berry,simon,aha,sam,wu-li,jordan,jackiw,resource,li-book},
particles with spin and magnetic moment moving in time-dependent
magnetic fields have received much attention
\cite{wang,wagh1,wagh2,fer,fer-pla,layton,gao,ni,zhang,zhu00,pra01,jpa01,jpa02,ni-book},
though the subject is rather old and some discussions can be found in
the textbook \cite{landau}. Neutral particles are of special interest
since the problem is easier and the Schr\"odinger equation can be
solved analytically in some special cases, say, uniform magnetic
fields with a fixed direction or rotating ones. Thus the model is
very suitable for the study of time evolution, cyclic solutions and
geometric phases etc. However, some problems in this model are still
not clear. First, the Schr\"odinger equation for general spin, or
even for spin $1/2$, in an arbitrarily varying magnetic field seems
impossible to be solved analytically. Second, though the existence of
cyclic solutions in a given time interval may be ensured by the
existence of eigenvectors for the unitary time evolution operator, it
seems not clear there are how many ones in the general case. Third,
for spin $1/2$ it is well known that any cyclic solution in an
arbitrary magnetic field has a nonadiabatic geometric phase
proportional to the solid angle subtended by the closed trace of the
spin vector. For higher spin, however, this is true only for cyclic
solutions with special initial conditions \cite{layton,gao,jpa02}.
For more general cyclic solutions in a rotating magnetic field, we
have shown that the nonadiabatic geometric phase contains an extra
term in addition to the one proportional to the solid angle. The
extra term vanishes automatically for spin $1/2$. For higher spin,
however, it depends on the initial condition \cite{jpa02}. It is
still not clear what is the relation between the nonadiabatic
geometric phase and the solid angle for general cyclic solutions in
an arbitrary magnetic field. In this paper we are going to deal with
these problems, and try to solve them to some extent. Besides the
theoretical interest in itself and other applications
\cite{li-book,zhu00}, this subject has been recently recognized to be
of great interest in the physics of quantum computation
\cite{wang01,zhu02}.

Consider a neutral particle with spin $s$ ($s=1/2, 1, 3/2, \ldots$)
and magnetic moment $\bm\mu=\mu\mathbf s/s$, where $\mathbf s$ is the
spin operator in the unit of $\hbar$, satisfying $[s_i,s_j] ={\mathrm
i}\epsilon_{ijk}s_k$ (for spin $1/2$, $\mathbf s=\bm\sigma/2$ and
$\bm\mu=\mu\bm\sigma$). In a uniform but time-dependent magnetic
field $\mathbf B(t)=B(t)\mathbf n(t)$ where $\mathbf n(t)$ is a unit
vector, it has the Hamiltonian $H(t)=-\bm\mu\cdot\mathbf
B(t)=-\hbar\omega_B(t) \mathbf s\cdot\mathbf n(t)$, where
$\omega_B(t)=\mu B(t)/s\hbar$, and the Schr\"odinger equation
$\mathrm i\hbar\partial_t\psi(t)=H(t)\psi(t)$ takes the form
\begin{equation}\label{1}
\partial_t\psi(t)=\mathrm i\omega_B(t)\mathbf s\cdot\mathbf n(t)\psi(t).
\end{equation}
Define the spin vector as
\begin{equation}\label{2}
{\mathbf v}(t)={\bm(}\psi(t), {\mathbf s}\psi(t){\bm )}.
\end{equation}
Using Eq. (\ref{1}) it is easy to show that it obeys the equation
\begin{equation}\label{3}
\dot {\mathbf v}(t)=-\omega_B(t){\mathbf n}(t)\times {\mathbf v}(t).
\end{equation}
We are not going to solve Eq. (\ref{1}) in the general case since
this seems impossible. However, we will show that the time evolution
operator for Eq. (\ref{1}) can be obtained without any chronological
product if one can find one nontrivial (nonzero) solution, say, a
unit vector $\mathbf e(t)$, to Eq. (\ref{3}). This is of interest
since the latter is easier and more cases can be solved
\cite{fer-pla}. Actually, Eq. (\ref{1}) involves operators while Eq.
(\ref{3}) involves only c-numbers. On the other hand, if transformed
to a matrix equation, Eq. (\ref{1}) involves $2s+1$ complex
variables, while Eq. (\ref{3}) involves only three real ones
(actually two since it is easy to see that $\mathbf v^2(t)=\mathbf
v^2(0)$). Using the time evolution operator and Eq. (\ref{3}), one
can discuss cyclic solutions and geometric phases in a most general
way. In particular, we will show that there exist at least $2s+1$
cyclic solutions in any time interval. A general relation between the
nonadiabatic geometric phase and the solid angle subtended by the
closed trace of the spin vector will be established.

To begin, we take an arbitrary unit vector $\mathbf e_0$, and the
eigenstate of ${\mathbf s} \cdot{\mathbf e}_0$ with eigenvalue $m$
will be denoted by $\chi_m$. We take the initial state of the system
to be $\psi(0)=\chi_m$, that is
\begin{equation}\label{4}
{\mathbf s}\cdot{\mathbf e}_0\psi(0)=m\psi(0),\quad
m=s,s-1,\ldots,-s.
\end{equation}
Obviously ${\mathbf v}(0)=m\mathbf e_0$ in this initial state. Now we
define a vector ${\mathbf e}(t)$ by Eq. (\ref{3}) with the initial
condition $\mathbf e_0$, that is
\begin{equation}\label{5}
\dot {\mathbf e}(t)=-\omega_B(t){\mathbf n}(t)\times {\mathbf e}(t)
\end{equation}
with ${\mathbf e}(0)={\mathbf e}_0$. We would assume that $\mathbf
B(t)$ varies continuously, so that any solution ${\mathbf e}(t)$ is
well behaved. As pointed out above, $\mathbf e^2(t)=\mathbf e_0^2$,
so ${\mathbf e}(t)$ is a unit vector at any time. We have proven in
Ref. \cite{jpa02} that
\begin{equation}\label{6}
{\mathbf s}\cdot{\mathbf e}(t)\psi(t)=m\psi(t)
\end{equation}
holds at all later times. To be self-contained, we repeat here the
proof by induction.

By definition, Eq. (\ref{6}) is valid at $t=0$. We assume that it is
valid at time $t$, what we need to do is to show that it is also true
at time $t+\Delta t$ where $\Delta t$ is an infinitesimal increment
of time. In fact, using Eqs. (\ref{1}) and (\ref{5}) we have
\begin{subequations}\label{7}
\begin{equation}\label{7a}
\psi(t+\Delta t)=\psi(t)+\mathrm i\omega_B(t){\mathbf s}\cdot{\mathbf
n}(t) \psi(t)\Delta t,
\end{equation}
\begin{equation}\label{7b}
{\mathbf e}(t+\Delta t)={\mathbf e}(t)-\omega_B(t){\mathbf
n}(t)\times{\mathbf e}(t) \Delta t.
\end{equation}
\end{subequations}
After some simple algebra, the conclusion is achieved.

Because ${\mathbf e}(t)$ is a unit vector, we can write in some
rectangular coordinates
\begin{equation}\label{8}
{\mathbf e}(t)={\bm (}\sin\theta(t)\cos\phi(t),\;
\sin\theta(t)\sin\phi(t),\; \cos\theta(t){\bm )}.
\end{equation}
Using the formula \cite{jpa02,wilcox}
\begin{equation}\label{9}
\mathrm e^{\mathrm i\xi{\mathbf s}\cdot{\mathbf b}}{\mathbf s}
\mathrm e^{-\mathrm i\xi{\mathbf s} \cdot{\mathbf b}}=[{\mathbf
s}-({\mathbf s}\cdot {\mathbf b}){\mathbf b}] \cos\xi+({\mathbf
b}\times{\mathbf s})\sin\xi+({\mathbf s}\cdot {\mathbf b}){\mathbf
b},
\end{equation}
where $\mathbf b$ is any unit vector, it is not difficult to show
that
\begin{equation}\label{10}
{\mathbf s}\cdot{\mathbf e}(t)=\mathrm e^{-\mathrm i\theta(t)\mathbf
s\cdot\mathbf d(t)} s_z \mathrm e^{\mathrm i\theta(t)\mathbf
s\cdot\mathbf d(t)},
\end{equation}
where
\begin{equation}\label{11}
\mathbf d(t)=\bm(-\sin\phi(t), \cos\phi(t), 0\bm).
\end{equation}
Therefore the eigenstate of ${\mathbf s}\cdot{\mathbf e}(t)$ with
eigenvalue $m$ is
\begin{equation}\label{12}
\psi(t)=\mathrm e^{\mathrm i\alpha_m(t)}\mathrm e^{-\mathrm
i\theta(t)\mathbf s\cdot\mathbf d(t)} \chi^0_{m},
\end{equation}
where $\chi^0_{m}$ is the eigenstate of $s_z$ with eigenvalue $m$,
and $\alpha_m(t)$ is a phase that cannot be determined by the
eigenvalue equation. However, $\alpha_m(t)$ is not arbitrary. To
satisfy the Schr\"odinger equation, it should be determined by the
other variables $\theta(t)$ and $\phi(t)$. In fact, the above
equation yields
\begin{equation}\label{13}
{\bm(}\psi(t),\psi(t+\Delta t){\bm)}=1+\mathrm i\dot\alpha_m(t)\Delta
t +\bm (\chi^0_{m}, \mathrm e^{\mathrm i\theta(t)\mathbf
s\cdot\mathbf d(t)}
\partial_t \mathrm e^{-\mathrm i\theta(t) \mathbf s\cdot\mathbf d(t)}
\chi^0_{m}\bm)\Delta t.
\end{equation}
Using the formula \cite{wilcox}
\begin{equation}\label{14}
\mathrm e^{-F(t)}\partial_t \mathrm e^{F(t)}=\int_0^1 \mathrm
e^{-\lambda F(t)}\dot F(t) e^{\lambda F(t)}\;\mathrm d\lambda,
\end{equation}
where $F(t)$ is any operator depending on $t$, and then using Eq.
(\ref{9}), we obtain
\begin{equation}\label{16}
{\bm(}\psi(t),\psi(t+\Delta t){\bm)}=1+\mathrm i\dot\alpha_m(t)\Delta
t +\mathrm i m[1-\cos\theta(t)]\dot\phi(t)\Delta t.
\end{equation}
On the other hand, from Eq. (\ref{1}) we have
\begin{equation}\label{17}
{\bm(}\psi(t),\psi(t+\Delta t){\bm)}=1+\mathrm i\omega_B(t){\mathbf
v}(t) \cdot{\mathbf n}(t)\Delta t.
\end{equation}
Note that $\mathbf v(t)$ and $\mathbf e(t)$ satisfy the same
equation, and $\mathbf v(0)=m\mathbf e_0$, we have $\mathbf
v(t)=m\mathbf e(t)$. Comparing the two results above and taking this
relation into account, we obtain
\begin{equation}\label{18}
\dot\alpha_m(t)=-m[1-\cos\theta(t)]\dot\phi(t)+m\omega_B(t){\mathbf
e}(t) \cdot{\mathbf n}(t).
\end{equation}
Therefore
\begin{equation}\label{19}
\alpha_m(t)-\alpha_m(0)=m\alpha(t),
\end{equation}
where
\begin{equation}\label{20}
\alpha(t)=-\int_0^t [1-\cos\theta(t')]\dot\phi(t')\;\mathrm dt' +
\int_0^t \omega_B(t'){\mathbf e}(t') \cdot{\mathbf n}(t') \;\mathrm
dt'.
\end{equation}
Substituting into Eq. (\ref{12}) we obtain
\begin{equation}\label{21}
\psi(t)=\mathrm e^{-\mathrm i\theta(t)\mathbf s\cdot\mathbf d(t)}
\mathrm e^{\mathrm i\alpha(t) s_z}\mathrm e^{\mathrm i\theta(0)
\mathbf s\cdot\mathbf d(0)}\psi(0).
\end{equation}
We denote the time evolution operator as $U(t)$, defined by the
equation $\psi(t)=U(t)\psi(0)$ with an arbitrary $\psi(0)$, then the
above equation is equivalent to
\begin{equation}\label{22}
U(t)\chi_m=\mathrm e^{-\mathrm i\theta(t)\mathbf s\cdot\mathbf d(t)}
\mathrm e^{\mathrm i\alpha(t) s_z}\mathrm e^{\mathrm i\theta(0)
\mathbf s\cdot\mathbf d(0)}\chi_m.
\end{equation}
Now an arbitrary initial state $\psi(0)$ can always be expanded as
\begin{equation}\label{23}
\psi(0)=\sum_m c_m \chi_m.
\end{equation}
Applying $U(t)$ to both sides of this equation, using Eq. (\ref{22}),
and noting that the operators on the right-hand side of that equation
is independent of $m$, we immediately realize that Eq. (\ref{21}) is
in fact valid for an arbitrary initial state. Thus we arrive at the
result
\begin{subequations}\label{24}
\begin{equation}\label{24a}
U(t)=\mathrm e^{-\mathrm i\theta(t)\mathbf s\cdot\mathbf d(t)}
\mathrm e^{\mathrm i\alpha(t) s_z}\mathrm e^{\mathrm i\theta(0)
\mathbf s\cdot\mathbf d(0)}.
\end{equation}
Using Eq. (\ref{9}), it can be recast in the form
\begin{equation}\label{24b}
U(t)=\mathrm e^{-\mathrm i\theta(t)\mathbf s\cdot\mathbf d(t)}
\mathrm e^{\mathrm i\theta(0) \mathbf s\cdot\mathbf d(0)} \mathrm
e^{\mathrm i\alpha(t) \mathbf s\cdot\mathbf e_0}.
\end{equation}
\end{subequations}
Eq. (\ref{24b}) is suitable for the general discussions below while
Eq. (\ref{24a}) may be more convenient for practical calculations.

Let us make some remarks on the result. First, we see that once a
solution of Eq. (\ref{5}) is found, the time evolution operator for
Eq. (\ref{1}) is available and it involves no chronological product.
The result depends formally on $\mathbf e_0$, but $\mathbf e_0$ is
merely an auxiliary object, hence the result must be essentially
independent of it, though it might be difficult to prove this
explicitly. In practical calculations, one should choose a solution
$\mathbf e(t)$ that is as simple as possible such that $U(t)$ can be
easily reduced to the simplest form. When this approach is used to
the simple cases such as rotating magnetic fields or ones with a
fixed direction, it indeed leads to the same results as those
obtained previously. Second, the operator $U(t)$ depends not only on
$\mathbf e(t)$, but also on the history of it. This is obvious from
Eq. (\ref{20}). Third, though $\phi(t)$ is indefinite when
$\theta(t)=0$ or $\pi$, the above result $U(t)$ is in fact well
behaved everywhere. That it is well defined at $\theta(t)=0$ is
obvious. If $\theta(0)=0$, there is no problem either. The case with
$\theta(0)=\pi$ can be avoided since one can always choose a
coordinate system such that $\theta(0)\ne\pi$. However, for a general
evolution, the case with $\theta(t)=\pi$ at some instant cannot be
avoided. Thus we must show that $U(t)$ is well behaved at
$\theta(t)=\pi$. Suppose that $\theta(t_0)=\pi$, then we have
$\phi(t_0^+)-\phi(t_0^-)=\pm\pi$, $\mathbf d(t_0^+)=- \mathbf
d(t_0^-)$, and $\alpha(t_0^+)-\alpha(t_0^-)=\mp 2\pi$. With these
relations it is not difficult to show that $U(t_0^+)=U(t_0^-)$. Since
both $U(t_0^+)$ and $U(t_0^-)$ are well defined, we may define
$U(t_0)=\lim_{t\to t_0}U(t)$. This makes $U(t)$ well defined and
continuous at $t=t_0$. Fourth, by straightforward calculations it can
be shown that $\partial_t U(t)=\mathrm i \omega_B(t) \mathbf s \cdot
\mathbf n(t) U(t)$ and $U(0)=1$, as expected.

Now we can go further to discuss cyclic solutions in any time
interval $[0,\tau]$ where $\tau$ is an arbitrarily given time. These
cyclic solutions are not necessarily cyclic in subsequent time
intervals with the same length, say, $[\tau,2\tau]$.

Since Eq. (\ref{5}) is a linear differential equation, the general
solution $\mathbf e(t)$ must depend on the initial vector $\mathbf
e_0$ linearly. Thus it can be written in a matrix form
\begin{equation}\label{25}
e_i(t)=E_{ij}(t)e_{0j},
\end{equation}
where the matrix $E(t)$ is obviously real. If both $\mathbf e_1(t)$
and $\mathbf e_2(t)$ are solutions to Eq. (\ref{5}), it is easy to
show that $\mathbf e_1(t)\cdot\mathbf e_2(t)=\mathbf
e_1(0)\cdot\mathbf e_2(0)$. Therefore the matrix $E(t)$ is an
orthogonal one, and its eigenvalues at any time $t$ has the form
$\{1,\sigma(t),\sigma^*(t)\}$, where $\sigma(t)$ is a complex number
with $|\sigma(t)|=1$, and $\sigma^*(t)$ its complex conjugate.

If $\sigma(\tau)\ne 1$, one eigenvector $\bm\eta(\tau)$ of the matrix
$E(\tau)$ with eigenvalue $1$ can be found, which satisfies
$E_{ij}(\tau) \eta_j(\tau) =\eta_i(\tau)$. It can be taken as real
and normalized. Now if we choose
\begin{equation}\label{26}
\mathbf e_0=\bm\eta(\tau),
\end{equation}
we have $e_i(\tau)=E_{ij}(\tau) e_{0j}=E_{ij}(\tau) \eta_j(\tau)=
\eta_i(\tau)=e_i(0)$, that is
\begin{equation}\label{27}
\mathbf e(\tau)=\mathbf e_0.
\end{equation}
This means that $\theta(\tau)=\theta(0)$ and $\mathbf d(\tau)=\mathbf
d (0)$, and leads to
\begin{equation}\label{28}
U(\tau)=\mathrm e^{\mathrm i\alpha(\tau)\mathbf s\cdot\mathbf e_0}.
\end{equation}
Now it is clear that with the initial condition $\psi(0)=\chi_m$ ($
m=s, s-1,\ldots, -s$), we have a cyclic solution in the time interval
$[0,\tau]$. More specifically, $\psi(\tau)=\mathrm e^{\mathrm
i\delta}\psi(0)$, where the total phase change is $\delta=m
\alpha(\tau),\mathrm{mod}~2\pi$, with $\alpha(\tau)$ given by
\begin{equation}\label{29}
\alpha(\tau)=-\Omega_e+\int_0^\tau \omega_B(t)\mathbf e(t)\cdot
\mathbf n(t)\;\mathrm d t,
\end{equation}
where
$$
\Omega_e=\int_0^\tau [1-\cos\theta(t)]\dot\phi(t)\,\mathrm d t
$$
is the solid angle subtended by the closed trace of $\mathbf
e(t)$. Notice that $\mathbf v(t)=m\mathbf e(t)$, the dynamic phase
$\beta=-\hbar^{-1}\int_0^\tau \langle H(t)\rangle\;\mathrm d t$ turns
out to be
\begin{equation}\label{30}
\beta=m\int_0^\tau \omega_B(t)\mathbf e(t)\cdot \mathbf n(t)\;\mathrm
d t.
\end{equation}
Therefore the nonadiabatic geometric phase is
\begin{equation}\label{31}
\gamma=\delta-\beta=-m\Omega_e,\quad \mathrm{mod}~2\pi.
\end{equation}
Since $\Omega_e=\epsilon(m)\Omega_v,\mathrm{mod}~4\pi$, where
$\Omega_v$ is the solid angle subtended by the closed trace of the
spin vector, we have finally
\begin{equation}\label{32}
\gamma=-|m|\Omega_v,\quad \mathrm{mod}~2\pi,
\end{equation}
in accord with the results previously obtained
\cite{layton,gao,jpa02}. Thus we see that $2s+1$ cyclic solutions are
available in any time interval $[0,\tau]$, and all phases can be
expressed in terms of the unit vector $\mathbf e(t)$.

States with initial condition other than the above ones are in
general not cyclic ones, even those in which $\mathbf v(0)$ points in
the direction of $\mathbf e_0$ such that $\mathbf v(\tau)=\mathbf
v(0)$. However, if $\alpha(\tau)/\pi$ happens to be a rational number
other than an even integer (the case with $\alpha(\tau)/\pi$ an even
integer will be discussed below), some other cyclic solutions may be
available. To be more specific, let $\alpha(\tau)=p\pi/n$, where $n$
is a natural number and $p$ an integer. When $n=1$, $p$ is an odd
number, and when $n>1$ it is prime with $p$. If $s\ge n$, we have
cyclic solutions with initial condition, say (no such solution exits
for $s=1/2$),
\begin{equation}\label{33}
\psi(0)=\sum_{j=-N_1}^{N_2} c_j\chi_{m+2nj},
\end{equation}
where $N_1$ and $N_2$ are nonnegative integers, and
$$
m-2nN_1\ge -s,\quad m+2nN_2\le s, \quad \sum_{j=-N_1}^{N_2}
|c_j|^2=1.
$$
In this initial state
\begin{equation}\label{a1}
\mathbf v(0)=v_0\mathbf e_0,
\end{equation}
where $v_0=\sum_{j=-N_1}^{N_2} (m+2nj) |c_j|^2$ may be either
positive or negative. It is easy to show that $\psi(\tau)=\mathrm
e^{\mathrm i\delta}\psi(0)$, where $\delta=m
\alpha(\tau),\mathrm{mod}~2\pi$. Because of Eq. (\ref{a1}), we have
$\mathbf v(t)=v_0\mathbf e(t)$, and the dynamic phase is
$$
\beta=-\hbar^{-1}\int_0^\tau \langle H(t)\rangle\;\mathrm d t
=\int_0^\tau \omega_B(t)\mathbf v(t)\cdot \mathbf n(t)\;\mathrm d t
=v_0\int_0^\tau \omega_B(t)\mathbf e(t)\cdot \mathbf n(t)\;\mathrm d
t.
$$
Using Eq. (\ref{29}), we have
\begin{equation}\label{a2}
\beta=v_0[\alpha(\tau)+\Omega_e].
\end{equation}
This holds regardless of the values of $v_0$ and $\alpha(\tau)$, as
long as Eq. (\ref{a1}) is valid. Suppose that in the process from
$t=0$ to $t=\tau$, $\mathbf e(t)$ encircles the polar axis $K$ times
($K>0$ for anticlockwise traces and $K<0$ for clockwise ones), then
we have
\begin{equation}\label{a3}
\Omega_e+\Omega_{-e}=4\pi K.
\end{equation}
This leads to $v_0\Omega_e=|v_0|\Omega_v+2\pi K(v_0-|v_0|)$. The
geometric phase turns out to be
\begin{equation}\label{34}
\gamma=-|v_0|\Omega_v+2\pi K (|v_0|-v_0)+(m-v_0)p\pi/n,\quad
\mathrm{mod}~2\pi.
\end{equation}
In this case we see that the geometric phase contains extra terms in
addition to the one proportional to $\Omega_v$, unless the sum of
these extra terms happens to be an integral multiple of $2\pi$. Note
that the above relation holds for the case $v_0=0$ as well, though
$\Omega_v$ is not well defined in this case.

If it happens that $\alpha(\tau)=2k\pi$, then $U(\tau)=\mathrm
e^{\mathrm i 2\pi ks}$ becomes a c-number and all solutions are
cyclic in the time interval $[0,\tau]$. However, this is true only on
the premise of (\ref{27}). Thus $\alpha(\tau)=2k\pi$ alone is not a
sufficient condition for all solutions to be cyclic, but it is easy
to show that it is a necessary one. A sufficient condition is
$\sigma(\tau)=1$.

Now if $\sigma(\tau)=1$, $E(\tau)$ becomes a unit matrix. In this
case any vector is its eigenvector with eigenvalue $1$. Therefore
Eqs. (\ref{27}) and (\ref{28}) hold for any unit vector $\mathbf
e_0$. In particular, we have
\begin{equation}\label{35}
U(\tau)=\mathrm e^{\mathrm i\alpha_1(\tau) s_x}= \mathrm e^{\mathrm
i\alpha_2(\tau) s_y} =\mathrm e^{\mathrm i\alpha_3 (\tau)s_z},
\end{equation}
where $\alpha_1(\tau)$ is given by Eq. (\ref{29}) with $\mathbf e_0
=\mathbf e_x$ and similarly for $\alpha_2(\tau)$ and $\alpha_3
(\tau)$. Since $s_x$, $s_y$ and $s_z$ are independent operators, the
above equation cannot hold unless $\alpha_i(\tau)=2\pi k_i$ where
$k_i$ are integers such that $U(\tau)$ becomes a c-number. In
general, with an initial unit vector $\mathbf e_0$, we have
\begin{equation}\label{36}
\alpha(\tau)=2\pi k,
\end{equation}
and
\begin{equation}\label{37}
U(\tau)=\mathrm e^{\mathrm i\alpha(\tau)\mathbf s\cdot\mathbf e_0}
=\mathrm e^{\mathrm i 2\pi ks}.
\end{equation}

Let us see what is the dependence of $\alpha(\tau)$ or $k$ on the
direction of $\mathbf e_0$. Consider two initial unit vectors
$\mathbf e_0$ and $\mathbf e'_0$, whose difference $\delta \mathbf
e_0=\mathbf e'_0-\mathbf e_0$ is infinitesimal (then $\delta \mathbf
e_0 \cdot \mathbf e_0=0$). The difference in $\alpha(\tau)$ is,
according to Eq. (\ref{29}),
\begin{equation}\label{38}
\delta\alpha(\tau)=\alpha'(\tau)-\alpha(\tau)
=-\delta\Omega_e+\int_0^\tau \omega_B(t)\delta\mathbf e(t)\cdot
\mathbf n(t)\;\mathrm d t.
\end{equation}
Since $e_i(t)=E_{ij}(t)e_{0j}$, we have $\delta e_i(t)=
E_{ij}(t)\delta e_{0j}$, and the second term in the above equation is
consequently infinitesimal. Moreover, the trace of $\mathbf e'(t)$ is
very close to that of $\mathbf e(t)$, thus the difference in the
solid angles subtended by them is infinitesimal. Therefore both
$\delta \alpha(\tau)$ and $\delta k=\delta \alpha(\tau)/2\pi$ are
infinitesimal as well. Now that $k$ can take only integer values, an
obvious consequence is that $\delta\alpha(\tau)=0$ and $k'=k$.

A subtle case has been overlooked in the above discussions, however.
This happens when $\mathbf e(t)$ goes by the south pole
($\theta=\pi$) on one side and $\mathbf e'(t)$ goes by it on the
other. In other words, the closed trace of $\mathbf e(t)$ as a whole
goes across the south pole when $\mathbf e_0$ varies to $\mathbf
e'_0$. In this case $\alpha(\tau)$ will be changed by an integral
multiple of $4\pi$ and $k$ by an even integer. One can of course
removing this change by rotating the coordinate system such that the
trace of $\mathbf e(t)$ does not go across the south pole. However,
this can only be done locally, and globally it is impossible in
general. In other words, one cannot choose a coordinate system such
that $k$ is the same for all $\mathbf e_0$, except for some simple
cases, say, magnetic fields with a fixed direction. The above
dependence of $\alpha(\tau)$ on $\mathbf e_0$ has no consequence on
the time evolution operator as expected. This is easily seen from Eq.
(\ref{37}): when $k$ changes by an even integer, $U(\tau)$ remains
the same.

In the special case where $k$ is the same for all $\mathbf e_0$, a
geometric explanation of $k$ is available. If we take the unit vector
$\mathbf e_0'=-\mathbf e_0$ as an initial condition to Eq. (\ref{5}),
then the solution is $\mathbf e'(t)=-\mathbf e(t)$. Since
$\alpha(\tau)$ corresponding to $\mathbf e(t)$ and $\alpha'(\tau)$
corresponding to $\mathbf e'(t)$ are equal as assumed, we have,
according to Eq. (\ref{29}), $2\alpha
(\tau)=-(\Omega_e+\Omega_{-e})$. As before,
$\Omega_e+\Omega_{-e}=4\pi K$, and $2\alpha(\tau)=4\pi k$, so that
$k=-K$. Here $K$ is the winding number of $\mathbf e(t)$ around the
polar axis. Unfortunately, the geometric meaning of $k$ in the
general case is still not clear.

Now that $U(\tau)$ is a c-number, all solutions to Eq. (\ref{1})
become cyclic in the time interval $[0,\tau]$. The total phase change
is $\delta=2\pi ks=s\alpha(\tau),\mathrm{mod}~2\pi$. Since this phase
is determined only up to an integral multiple of $2\pi$, the
dependence of $\alpha(\tau)$ or $k$ on the initial vector $\mathbf
e_0$ does not affect the result. For convenience we take the
$\alpha(\tau)$ or $k$ that is the same as that appears below. If in
the initial state $\mathbf v(0)\ne 0$, we take $\mathbf e_0=\mathbf
v(0)/v_0$ where $v_0=|\mathbf v(0)|>0$. Thus $\mathbf v(t)=v_0
\mathbf e(t)$ and $\Omega_e=\Omega_v$. The dynamic phase is, as shown
in Eq. (\ref{a2}), $\beta= v_0 [\alpha(\tau)+\Omega_v]$. It should be
remarked that both $\alpha(\tau)$ and $\Omega_v$ may be different by
an integral multiple of $4\pi$ in different coordinate systems.
However, $\beta$ has a definite value, independent of the choice of
coordinate systems, which can be easily seen from the expression
$\beta=v_0\int_0^\tau \omega_B(t)\mathbf e(t)\cdot \mathbf
n(t)\;\mathrm d t$. The geometric phase turns out to be
\begin{equation}\label{40}
\gamma=-v_0 \Omega_v+(s-v_0)2\pi k,\quad \mathrm{mod}~2\pi.
\end{equation}
Here the first term is the familiar one, but an extra term appears.
If $s=1/2$, we have $v_0=1/2$ for any initial state, and the above
result reduces to $\gamma=-\Omega_v/2$, a well-known result. For
higher spin, however, the extra term depends on the initial state, as
both $v_0$ and $k$ depend on it. It vanishes (mod $2\pi$ of course)
when $s-v_0$ is an integer, especially when the initial state is an
eigenstate of $\mathbf s\cdot \mathbf e_0$ (it cannot be an
eigenstate of $\mathbf s\cdot \mathbf e_0'$ with some other unit
vector $\mathbf e_0'$ since $\mathbf v(0)$ points in the direction of
$\mathbf e_0$), as expected. If in the initial state $\mathbf
v(0)=0$, we have $\mathbf v(t)=0$ and thus $\beta=0$. It is easy to
see that Eq. (\ref{40}) holds in this case as well, though $\Omega_v$
is not well defined.

The general result (\ref{40}) has been confirmed by practical
calculations in the case of a rotating magnetic field, where both
$\gamma$ and $\Omega_v$ can be calculated explicitly \cite{jpa02}.
For a magnetic field with a fixed direction, it is easier to carry
out similar calculations to verify this result.

In summary, we have shown that the time evolution operator for the
Schr\"odinger equation (\ref{1}) can be obtained if one nontrivial
solution to Eq. (\ref{5}) can be found. We proved that at least
$2s+1$ cyclic solutions of the Schr\"odinger equation exist in any
time interval. These cyclic solutions can be worked out in principle
if the general solution to Eq. (\ref{5}) is known. There may exist
some particular time interval in which all solutions are cyclic. The
nonadiabatic geometric phase for cyclic solutions contains in general
extra terms in addition to the familiar one that is proportional to
the solid angle subtended by the trace of the spin vector. For spin
$1/2$ there is no such extra term.

\begin{acknowledgments}
This work was supported by the National Natural Science Foundation of
China (10275098), and by the Foundation of the Advanced Research
Center of Sun Yat-Sen University (02P3).
\end{acknowledgments}


\end{document}